\documentclass[prd,12pt]{article}
\pdfoutput=1 

\usepackage{diagbox}

\usepackage{tikz}
\usepackage{amsmath,amssymb,graphicx,multirow,xspace,slashed,array,booktabs}
\usepackage[colorlinks=true,urlcolor=blue,anchorcolor=blue,citecolor=blue,filecolor=blue,linkcolor=blue,menucolor=blue,pagecolor=blue]{hyperref}
\usepackage[compress,numbers]{natbib}
\usepackage{placeins}
\usepackage{subcaption}
\usepackage{booktabs}
\usepackage{blindtext, rotating}
\usepackage{afterpage}
\usepackage{enumitem}
\usepackage{ marvosym }
\usepackage{authblk} 
\usepackage{verbatim}
\usepackage{soul} 
\usepackage[normalem]{ulem}
\usepackage{pifont}
\usepackage{booktabs}
\usepackage{bm}
\usepackage{cleveref}
\usepackage{siunitx}
\usepackage{tikz-feynman}
\tikzfeynmanset{compat=1.1.0}
\usepackage{cases}

\usepackage{floatrow}
\newfloatcommand{capbtabbox}{table}[][\FBwidth]

\usepackage[font=footnotesize,labelfont=bf]{caption}

\usepackage{lineno}

\allowdisplaybreaks

\long\def\symbolfootnote[#1]#2{\begingroup%
\def\thefootnote{\fnsymbol{footnote}}\footnote[#1]{#2}\endgroup}

\allowdisplaybreaks

\makeatletter
	
	\@addtoreset{equation}{section}
\makeatother

\newcommand{\newc}{\newcommand}
\newc{\gsim}{\lower.7ex\hbox{$\;\stackrel{\textstyle>}{\sim}\;$}}
\newc{\lsim}{\lower.7ex\hbox{$\;\stackrel{\textstyle<}{\sim}\;$}}
\newc{\gev}{\,{\rm GeV}}
\newc{\mev}{\,{\rm MeV}}
\newc{\ev}{\,{\rm eV}}
\newc{\kev}{\,{\rm keV}}
\newc{\tev}{\,{\rm TeV}}
\newc{\MHT}{$H_T^{\text{miss}}$}
\newc{\MET}{$\slashed{E}_T$}
\newc{\MTT}{$M_{T2}$}

\def\ln{\mathop{\rm ln}}

\newc{\mz}{M_Z}
\newc{\mpl}{M_*}
\newc{\mw}{m_{\rm weak}}
\newc{\nr}[1]{N^c_R{}_{#1}}

\usepackage{accents}
\newlength{\dhatheight}

\def\beq{\begin{equation}}
\def\eeq{\end{equation}}
\newcommand{\bea}{\begin{eqnarray}\begin{aligned}}
\newcommand{\eea}{\end{aligned}\end{eqnarray}}
\def\bitem{\begin{itemize}}
\def\eitem{\end{itemize}}

\begin{document}
\baselineskip 0.6cm

\begin{titlepage}

\vspace*{-0.5cm}

\thispagestyle{empty}

\begin{center}

\vskip 1cm

{{\Huge
Note on Warped Compactification \\[1ex]}
{\Large  -- Finite Brane Potentials and Non-Hermiticity --}
}

\vskip 1cm

\vskip 1.0cm
{\large Sudhakantha Girmohanta$^{1,2}$, Yuichiro Nakai$^{1,2}$,  Motoo Suzuki$^{3,4}$, \\[1ex]
Yaoduo Wang$^{1,2}$ and Junxuan Xu$^{1,2}$}
\vskip 0.5cm
{\it
$^1$Tsung-Dao Lee Institute, Shanghai Jiao Tong University, \\
No.~1 Lisuo Road, Pudong New Area, Shanghai, 201210, China \\
$^2$School of Physics and Astronomy, Shanghai Jiao Tong University, \\
800 Dongchuan Road, Shanghai, 200240, China \\
$^3$Department of Physics, Harvard University, Cambridge, MA 02138, USA \\
$^4$Institute of Particle and Nuclear Studies,\\
	        High Energy Accelerator Research Organization (KEK), Tsukuba, 305-0801, Japan}
\vskip 1.0cm

\end{center}

\begin{abstract}

We study radius stabilization in the Randall-Sundrum model without assuming any unnaturally large stabilizing scalar potential parameter at the boundary branes ($\gamma$) by the frequently used superpotential method.
Employing a perturbative expansion in $1/\gamma^2$ and the backreaction parameter, we obtain approximate analytical expressions for the radion mass and wavefunction. We validate them through a dedicated numerical analysis, which solves the linearized coupled scalar and metric field equations exactly. It is observed that the radion mass decreases with decreasing $\gamma$. Below a critical value of $\gamma$, the radion becomes tachyonic, suggesting destabilization of the extra dimension. We also address the issue of non-Hermiticity of the differential operator that determines the radion and Kaluza-Klein (KK) mode wavefunctions in the finite $\gamma$ limit. It is accomplished by finding an explicit form of the general scalar product that re-establishes the orthogonality in the KK decomposition.

\end{abstract}

\flushbottom

\end{titlepage}

\section{Introduction}\label{intro}

Randall and Sundrum (RS) realized a 5D model of spacetime whose extra dimension is anti-dS (AdS) and bounded by two 3-branes called the UV and IR branes~\cite{Randall:1999ee}. In this model, although the extra dimension itself is not too large as compared to the Planck length, the physical mass scale on the IR brane is red-shifted by the exponential warp factor. A massless mode is associated with the fluctuation of the radius of the extra dimension, called the `radion', that must receive a mass to be consistent phenomenologically. This is achieved by a radius stabilization mechanism as pioneered by Goldberger and Wise (GW)~\cite{Goldberger:1999uk, Goldberger:1999un}. The AdS/CFT correspondence~\cite{Maldacena:1997re,Gubser:1998bc,Witten:1998qj} implies that this model is dual to a 4D nearly conformal field theory for a large number of colors $N$~\cite{Arkani-Hamed:2000ijo,Rattazzi:2000hs}, where moving along the extra dimension is mapped to the renormalization group evolution in the energy scale. The UV brane represents a UV cut-off, while the stabilized IR brane signifies a spontaneous breakdown of the conformal invariance, resulting in a pseudo-Nambu Goldstone boson (pNGB) dilaton, that is interpreted as the dual to the radion. 

The RS model has recently received a renewed interest as it leads to a supercooled phase transition in the early Universe
that generates a stochastic gravitational wave background detectable by ongoing and near-future experiments.
The authors of ref.~\cite{Creminelli:2001th} argued that the high-temperature system is described by
the AdS-Schwarzschild (AdS-S) spacetime with the IR brane replaced by an event horizon.
As the temperature decreases, the system experiences a phase transition from the AdS-S to RS spacetime,
which corresponds to the confinement-deconfinement phase transition in the dual CFT picture.
The phase transition is generally strongly first-order and driven through a supercooling phase. 
The associated gravitational wave production emerges as a key cosmological probe of the model, where it has been analyzed by using the radion effective potential~\cite{Creminelli:2001th,Randall:2006py,vonHarling:2017yew, Baratella:2018pxi,Fujikura:2019oyi,Girmohanta:2023sjv,Fujikura:2023lkn, Mishra:2023kiu, Koutroulis:2024wjl,Agashe_Du_Ekhterachian_Kumar_Sundrum_2020,Agashe_Du_Ekhterachian_Kumar_Sundrum_2021}.
Ref.~\cite{Mishra:2024ehr} recently considered the back reaction  effect of a modified GW scalar potential.
In their radius stabilizing mechanism, GW used a weak-field limit, where the backreaction of the stabilizing scalar ($\phi$) is negligible. This approximation aids in the extraction of the radion effective potential. On the other hand, a `superpotential' method~\cite{Csaki:1999mp} can be used to solve the Einstein equation and $\phi$ equation of motion simultaneously, thereby automatically taking care of the backreaction. However, it is difficult to extract the radion effective potential and analyze the phase transition from this method for a fixed brane and bulk potential for $\phi$, because changing the brane separation induces a change in the brane potentials in the superpotential approach. 
Ref.~\cite{Konstandin:2010cd} has made progress with detuning the brane potentials with the backreaction partially taken account from the scalar.
Yet, in the previous studies, a limit was always assumed for a brane potential parameter called $\gamma$ which effectively fixes the $\phi$ vacuum expectation value on both of the branes by hand. Here, we will study the superpotential method without this approximation. This will be valuable for analyzing the parameter space of the phase transition in its full generality.

In order to take into account the finite-$\gamma$ effects, in the present note, we will develop a series expansion, in powers of both the backreaction parameter and the aforementioned $\gamma$, and elucidate how a finite $\gamma$ modifies the radion mass and wavefunction profile. To supplement our analysis further, we will numerically solve the coupled partial differential equations exactly and obtain the radion mass as a function of the parameters of the superpotential. Interestingly, we find that the radion mass decreases when one decreases $\gamma$, and the radion becomes tachyonic below a certain threshold $\gamma$, suggesting a possible breakdown of the stabilization.
~\footnote{
The radion becomes tachyonic in scenarios with two de Sitter branes without the GW field~\cite{Chacko_Fox_2001} (and related works \cite{Karch_Randall_2020, Mishra_2023}), whereas our setup involves Minkowski branes with the GW field.
}. 
Therefore, a minimum $\gamma$ is required to stabilize the extra dimension.  Our results on the radion mass can have important cosmological consequences including the phase transition for the general parameters

We will also deal with the issue of non-Hermiticity of the operator that determines the radion and Kaluza-Klein (KK) mode wavefunctions and mass eigenvalues in the finite $\gamma$ case. This arises due to the finite strip of extra dimension and non-trivial boundary conditions~\cite{Csaki:2000zn}. To illustrate, consider the operator $d^2/dy^2$, where $y$ denotes the coordinate along the extra dimension. This operator is not Hermitian with finite extra dimension and general boundary conditions. This is because one picks up residual boundary terms in the usual scalar product that is not guaranteed to be zero for a generic boundary condition, and we will see that this is indeed the case applicable to the superpotential method. Nevertheless, one needs a proper scalar product such that, in the KK decomposition, one can use the usual orthogonality and completeness relation of the wavefunctions. In order to solve this problem, we will define explicitly the general scalar product that restores the Hermiticity and results in the canonical kinetic terms for the KK mode expansion, thereby establishing the consistency of the method. 

The rest of the paper is organized as follows. In section~\ref{model}, we review the superpotential method and derive the form of the general scalar product. Section~\ref{perturbative} discusses the perturbative procedure for obtaining the radion mass and wavefunction. Section~\ref{numerical} contains details of our numerical analysis and results for the radion mass. Finally, we conclude in section~\ref{conclusion}.
Some details of calculations are summarized in appendices.

\section{The RS model with finite brane potentials}\label{model}

We begin with a review of the GW mechanism for radion stabilization in the RS model,
where the background scalar and metric equations are solved with the superpotential method~\cite{Csaki:2000zn}
(see also ref.~\cite{Lee:2021wau} for the case of multi-brane models).
For scalar fluctuations around the background configuration,
we see that the linearized field equations and boundary conditions are reduced to
a Schr\"{o}dinger-like equation.
It was found that the orthogonality of eigenfunctions for the Schr\"{o}dinger-like equation
is lost with the naive scalar product in the case of finite brane potentials.
We will point out that the Hermitian property is recovered by defining a new scalar product of eigenfunctions.

\subsection{Brief review of the superpotential method}

We consider the 5D RS model, namely, an $S^1/Z_2$ orbifold with two 3-branes placed
at the fixed points of $y=0, \pi R$ where $y \equiv x^4$ is the coordinate of the extra dimension and $R$ is the radius,
and assume the GW mechanism to stabilize the radius.
The bulk action with a 5D scalar field $\phi$ is given by
\begin{equation}
\label{bulkaction}
    S = -M_5^3\int d^5 x\sqrt{g} \, \mathcal{R} \, +\int d^5x\sqrt{g}\left(\frac{1}{2}\nabla\phi\nabla\phi-V(\phi)\right)
    \, - \, \int d^4x\sqrt{g^{P}_4}\lambda_{P}(\phi) \, - \, \int d^4x\sqrt{g^{T}_4}\lambda_T(\phi) \, ,
\end{equation}
where $M_5$ denotes the 5D Planck scale, $\mathcal{R} = g^{MN}R_{MN}$ ($M = \mu,4$) is the Ricci scalar,
$g$ is the bulk metric, $V$ is a bulk scalar potential,
$g^{P, \, T}_4$ are the induced metrics on the branes at $y_{P, \, T}=0, \pi R$, respectively,
and $\lambda_{P, \, T}$ are brane-localized potentials.
We first find the background configuration of the scalar field and the metric for this action.
The most general form of the configuration preserving the 4D Lorentz invariance is
\begin{equation}
\begin{split}
\label{ansatz}
    &\phi(x,y)=\phi_0(y) \, , \\[1ex]
    &ds^2 = e^{-2A(y)}\eta_{\mu\nu}dx^\mu dx^\nu-dy^2 \, .
\end{split}
\end{equation}
The Einstein equations are
\begin{equation}
    \mathcal{R}_{MN} =  \kappa^2\left(T_{MN}-\frac{1}{3}g_{MN}g^{AB}T_{AB}\right).
\end{equation}
Here, $\kappa^2 \equiv 1/(2M_5^3)$ and $T_{MN}$ denotes the energy-momentum tensor.
Inserting the ansatz of Eq.~\eqref{ansatz} into the Einstein equations and the equation of motion for $\phi$
derived from the action \eqref{bulkaction},
the scalar and metric field equations are obtained as
\begin{align}
    4{A'}^2-A'' &= -\frac{2\kappa^2}{3}V(\phi_0)-\frac{\kappa^2}{3}\sum_{i =P, \, T} \lambda_i(\phi_0)\delta(y-y_i) \, ,\label{2-5}\\[1ex]
    {A'}^2 &= \frac{\kappa^2{\phi_0'}^2}{12}-\frac{\kappa^2}{6}V(\phi_0) \, , \label{2-6}\\[1ex]
    \label{2-7}\phi_0'' &= 4A'\phi_0'+\frac{\partial V(\phi_0)}{\partial\phi}+\sum_{i =P, \, T}\frac{\partial\lambda_i(\phi_0)}{\partial\phi}\delta(y-y_i)  \, ,
\end{align}
where the prime $'$ denotes $\partial/\partial_y$ and $\partial_\mu$ is the derivative with respect to the comoving 4D spacetime coordinates $x^\mu$.

The boundary conditions for $A$ and $\phi_0$ are derived from matching the singular terms in the scalar and
metric field equations. 
To be specific, let us look at Eq.~\eqref{2-5}. Since $A,\phi_0$ are both physical quantities, they must be continuous.
On the other hand, the derivative of the metric is not required to be continuous and thus there can be a jump
in $A'$ at the branes, which means that the expression of $A'$ can have a term of the form $(A'(0+\epsilon)-A'(0-\epsilon))\theta(y)$ or $(A'(\pi R+\epsilon)-A'(\pi R-\epsilon))\theta(y-\pi R)$,
where $\epsilon \rightarrow +0$ and $\theta(y)$ is the Heaviside function. Then, $A''$ may contain the term $(A'(0+\epsilon)-A'(0-\epsilon))\delta(y)$ or $(A'(\pi R+\epsilon)-A'(\pi R-\epsilon))\delta(y-\pi R)$,
which can be used to match the $\delta$-function terms in Eq.~\eqref{2-5}, leading to
\begin{equation}
    \left[A'\right]|_i=\frac{\kappa^2}{3}\lambda_i(\phi_0)\label{BCA} \, ,
\end{equation}
where $\left[A'\right]\equiv A'(0+\epsilon)-A'(0-\epsilon) |_{\epsilon \rightarrow +0}$.
Similarly, Eq.~\eqref{2-7} gives
\begin{equation}
    \left[\phi_0'\right]|_i=\frac{\partial\lambda_i(\phi_0)}{\partial\phi}\label{BCphi} \, .
\end{equation}
We often call these boundary conditions as jump equations.

To get the analytic solution, we use the superpotential approach presented in ref.~\cite{superpotential}.
The form of the bulk scalar potential is taken as
\begin{equation}
    V(\phi) = \frac{1}{8}\left(\frac{\partial W(\phi)}{\partial \phi}\right)^2-\frac{\kappa^2}{6}W(\phi)^2\label{V} \, ,
\end{equation}
with $W(\phi)$ in the following form:
\begin{equation}
    W(\phi) = \frac{6k}{\kappa^2}-u \phi^2\label{W} \, .
\end{equation}
Here, $k$ is defined through the 4D Planck scale, $M_{\rm Pl}^2 \equiv (1-e^{-2k \pi R})/(k \kappa^2)$, and $u$ is a constant with mass dimension one. 
The brane-localized potentials are 
\begin{equation}
\begin{split}
    \lambda_{P}(\phi) &= W(\phi_P)+ \frac{\partial W}{\partial \phi} \biggr|_{\phi=\phi_P}(\phi-\phi_P)+\gamma^2_+ (\phi-\phi_P)^2 \, , \\[1ex]
    \label{branepot}\lambda_{T}(\phi) &= -W(\phi_T)- \frac{\partial W}{\partial \phi} \biggr|_{\phi=\phi_T}(\phi-\phi_T)+\gamma^2_- (\phi-\phi_T)^2\ ,
\end{split}
\end{equation}
where $\phi_{P, \, T}$ are values of $\phi$ at $y_{P, \, T}=0, \pi R$ respectively, and
$\gamma_{+,-}$ are parameters with mass dimension one.
For simplicity, $\gamma_{+}=\gamma_-=\gamma$ is assumed in the following analysis.
With the potential of Eq.~\eqref{V}, we can find that the solution to the first-order differential equations,
\begin{equation}
    \phi' = \frac{1}{2}\frac{\partial W(\phi)}{\partial \phi}\, , 
    \qquad A'=\frac{\kappa^2}{6}W(\phi) \, ,
    \label{sol}
\end{equation} 
is also the solution to the field equations satisfying the boundary conditions of Eqs.~\eqref{BCA}, \eqref{BCphi}.
Substituting Eqs.~\eqref{W}, \eqref{branepot} into Eq.~\eqref{sol} with the boundary conditions,
the solution of the background configuration is given by
\begin{align}
    \label{phi0} \phi_0(y) &= \phi_P e^{-uy}\, , \\[1ex]
    \label{A} A(y) &= ky +\frac{\kappa^2\phi_P^2}{12}e^{-2uy}\, .
\end{align}
We can see that the second term of Eq.~\eqref{A} gives the backreaction of the stabilizing field $\phi$ to the metric,
which will be important for the radion mass.
The value of $\phi_0$ at $y= r_0 \equiv \pi R$ helps to determine the distance between two branes:
\begin{equation}
    \phi_0(r_0)=\phi_T=\phi_P e^{-u r_0}\, ,
\end{equation}
which is equivalent to
\begin{equation}
     r_0 = \frac{1}{u} \ln\left(\frac{\phi_P}{\phi_T}\right)\, .
\end{equation}
The radius of the extra dimension is properly stabilized. 

Let us now consider perturbations around the background configuration.
We focus on spin-0 fluctuations,
\begin{align}
    &\phi(x,y) = \phi_0(y)+\varphi(x,y)\, , \\[1ex]
    &ds^2 = e^{-2A-2F\left(x,y\right)}\eta_{\mu\nu}dx^\mu dx^\nu-\left(1+G\left(x,y\right)\right)^2 dy^2\, ,
\end{align}
where $\varphi (x,y)$, $F(x,y)$ and $G(x,y)$ are the perturbations.
We linearize the Einstein equations and the scalar field equation of motion
to derive the equations for $F,G$ and $\varphi$.
The linearized Einstein equations are
\begin{equation}
    \delta \mathcal{R}_{MN} = \kappa^2\delta\tilde{T}_{MN}\label{Einsteineq}\, ,
\end{equation}
where $\tilde{T}_{MN}$ is defined as $\tilde{T}_{MN}=T_{MN}-\frac{1}{3}g_{MN}g^{AB}T_{AB}$.
The linearized Ricci tensor is given by
\begin{align}
    \delta \mathcal{R}_{\mu\nu} &= \eta_{\mu\nu}\square F+e^{-2A}\eta_{\mu\nu}\left(-F''+2A''(F+G)+A'(8F'+G')-8{A'}^2(F+G)\right)\notag\\ &+2\partial_\mu\partial_\nu F-\partial_\mu\partial_\nu G\, ,\\[1ex]
    \delta \mathcal{R}_{\mu5} &= 3\partial_\mu F'-3 A'\partial_\mu G\, ,\\[1ex]
    \delta \mathcal{R}_{55} &= e^{2A}\square G +4F''-8A'F'-4A'G'\, ,
\end{align}
and the energy-momentum tensor is
 \begin{align}
    \delta\tilde{T}_{\mu\nu} =& -\frac{2}{3}e^{-2A}\eta_{\mu\nu}\left(V'(\phi_0)\varphi-2V(\phi_0)F\right) \notag\\ &-\frac{1}{3}e^{-2A} \eta_{\mu\nu}\sum_i\left(\frac{\partial\lambda_i(\phi_0)}{\partial \phi}\varphi-2\lambda_i(\phi_0)(F+G)\right)\delta(y-y_i)\, ,\\[1ex]
    \delta\tilde{T}_{\mu5} =& \, \phi_0'\partial_\mu\varphi\, ,\\[1ex]
    \delta\tilde{T}_{55} =& \, 2\phi_0'\varphi_0'+\frac{2}{3}V'(\phi_0)\varphi+\frac{4}{3}V(\phi_0)G \notag\\ & +\frac{4}{3}\sum_i\left(\frac{\partial\lambda_i(\phi_0)}{\partial\phi}\varphi+\lambda_i(\phi_0)G\right)\delta(y-y_i)\, .
 \end{align}
We can see that for the $\mu\nu$ components, all terms except the terms
$2\partial_\mu\partial_\nu F-\partial_\mu\partial_\nu G$ in $\delta \mathcal{R}_{\mu\nu}$ and $\delta \tilde{T}_{\mu\nu}$ are proportional to the 4D Minkowski metric. Thus we have the relation,
\begin{equation}
    2\partial_\mu\partial_\nu F=\partial_\mu\partial_\nu G \, .
\end{equation}
Since when either $F$ or $G$ goes to zero, we should get the background configuration, the integration constant in the above relation should be zero. Therefore, we have $G=2F$ and now the radion is only described by the function $F(x,y)$.
Replacing $G$ with $2F$ in the above equations, we derive the coupled bulk equations for $F$ and $\varphi$. 

First we can integrate the equations for the $\mu 5$ components since both sides are total derivatives,
\begin{equation}
    \phi_0'\varphi = \frac{3}{\kappa^2}(F'-2A'F)\label{bulkeq1}\, ,
\end{equation}
where the ``integration constant" is taken to be zero so that the zero fluctuation limit is consistently obtained.
This equation says that $\varphi$ is totally determined by $F$.
Then we use the combination $e^{2A}\delta \mathcal{R}_{\mu\nu}+\delta \mathcal{R}_{55}$ to eliminate the $\varphi$ terms and get the equation that only includes $F$ and $\varphi'$:
\begin{equation}
    e^{2A}\square F+F''-2A'F'=\frac{2\kappa^2}{3}\phi_0'\varphi'\, \label{2.29}.
\end{equation}
Substituting Eq.~\eqref{bulkeq1} into this equation, we obtain the bulk equation only for $F$:
\begin{equation}
    F''-2A'F'-4A''F-2\frac{\phi_0''}{\phi_0'}F'+4A'\frac{\phi_0''}{\phi_0'}F = e^{2A}\square F\label{bulkeq2}\, .
\end{equation}
So far, we have not used the linearized scalar field equation derived from the action,
\begin{align}
    e^{2A}\square \varphi -\varphi''+4A'\varphi'+\frac{\partial^2V(\phi_0)}{\partial\phi^2}\varphi =& -\sum_i\left(\frac{\partial^2\lambda_i(\phi_0)\varphi}{\partial\phi^2}\varphi+2\frac{\partial\lambda_i(\phi_0)}{\partial\phi}F\right)\delta(y-y_i)\notag\\ &-6\phi_0'F'-4\frac{\partial V}{\partial\phi}F\, .
\end{align}
It is not hard to verify that this equation of motion can be derived from the combination of Eqs.~\eqref{2.29} and~\eqref{bulkeq2}, together with the background field equations~\eqref{2-5}, \eqref{2-6}, \eqref{2-7}.

As in the case of the background field solution,
we can find the boundary conditions by matching the singular second-derivative and delta-function terms.
The $\mu\nu$ and $55$ components of the Einstein equations give two boundary conditions
which are equivalent to each other while the scalar field equation leads to one condition,
\begin{align}
    \left[F'\right]|_i &= \frac{2\kappa^2}{3}\lambda_i(\phi_0)F+\frac{\kappa^2}{3}\frac{\partial\lambda_i (\phi_0)}{\partial\phi}\varphi\label{varjp1}\, ,\\[1ex]
     \left[\varphi'\right]|_i &= \frac{\partial^2 \lambda_i (\phi_0)}{\partial \phi^2}\varphi + 2\frac{\partial\lambda_i (\phi_0)}{\partial\phi}F\label{varjp2}\, .
\end{align}
However, we can find that the first condition is trivially satisfied by using Eq.~\eqref{bulkeq1}
and the boundary conditions for the background configuration.
Thus Eq.~\eqref{varjp2} is the only independent condition.
In summary, we have two bulk equations~\eqref{bulkeq1}, \eqref{bulkeq2} for $F$ and $\varphi$ and one boundary condition~\eqref{varjp2} at both branes, thus two constraints. 

The bulk equation for the scalar fluctuation $F$ in Eq.~\eqref{bulkeq2} can be rewritten
into a Schr\"{o}dinger-like equation.
In Eq.~\eqref{bulkeq2}, we change the variable by $dz e^{-A(z)}=dy$ and rescale the scalar field $\tilde{F}=e^{-\frac{3}{2}A}\frac{1}{\phi_{0,z}}F$ (the derivative with respect to $z$ is denoted by ``$_{,z}$"), which leads to the form,
\begin{align}
    &\hat H \tilde F=m^2 \tilde F
    \label{Schr}\ ,
\end{align}
where $\hat H$ is the  ``Hamiltonian" of the Schr\"{o}dinger-like equation,
\begin{align}
    &  \hat{H}=-\frac{d^2}{dz^2}+\hat{\lambda}(z)\ , \label{hamiltonian} \\
    & \hat{\lambda}(z)=\frac{9}{4}\left(A_{,z}\right)^2+\frac{5}{2}A_{,zz}-A_{,z}\frac{\phi_{0,zz}}{\phi_{0,z}}+2\left(\frac{\phi_{0,zz}}{\phi_{0,z}}\right)^2-\frac{\phi_{0,zzz}}{\phi_{0,z}}\label{lambda}\ .
\end{align}
To solve the equation, we perform the KK decomposition as usual,
\begin{align}
\tilde F=\sum_n \tilde F_n(z) f_n(x)\ , \label{eigenfunctions}
\end{align}
where the label $n$ denotes the eigenfunction with the eigenvalue $m_n^2$,
and $\tilde F_n(z)$ should form a complete and orthogonal set. 

\subsection{Non-Hermiticity}\label{non-hermiticity_section}
If the differential operator of the Schr{\"o}dinger-like equation is Hermitian,
the orthogonality of eigenfunctions and the completeness relation are automatically satisfied.
As we will see below, however, with finite brane-localized potentials of Eq.~\eqref{branepot},
a naive scalar product does not give the Hermitian property due to boundary contributions,
which obscures the KK decomposition. 
Here, we solve this issue by defining an improved scalar product to recover the Hermiticity.

Let us consider a naive scalar product,
\begin{align}
  (A,B)\equiv \int_{z_{P}}^{z_{T}} dz\, A^*\, B\ ,
\end{align}
where the interval of the integration is $z_{P} \leq z \leq z_{T}$ with $z_{P, \, T}$ corresponding to $y=0,~\pi R$, respectively. 
For the ``Hamiltonian" of Eqs.~\eqref{hamiltonian}, \eqref{lambda}, this scalar product leads to
\begin{align}
    \left(\hat{H}\tilde{F},\tilde{G}\right)
    &=\left(\tilde{F},\hat{H}\tilde{G}\right)+2\left[\tilde{F}^*_{,z} \, \tilde{G} \bigl|_{z_{P}} -\tilde{F}^*_{,z} \, \tilde{G} \bigl|_{z_{T}}\right.
    \left.-\tilde{G}_{,z} \, \tilde{F}^* \bigl|_{z_{P}}+\tilde{G}_{,z} \, \tilde{F}^* \bigl|_{z_{T}}
    \right] .
\end{align}
Here, $\tilde F(z)$ and $\tilde G(z)$ denote eigenfunctions defined in Eq.~\eqref{eigenfunctions}.
This relation indicates that, in general, the Hamiltonian is not Hermitian, $i.e.$ $\left(\hat{H}\tilde{F},\tilde{G}\right)\neq \left(\tilde{F},\hat{H}\tilde{G}\right)$ because 
the extra terms come from the boundaries. We may rewrite the relation as
\begin{align}
    (m^{2*}_F-m^2_G)\left(\tilde{F},\tilde{G}\right)
    &=2\left[\tilde{F}^*_{,z} \, \tilde{G} \bigl|_{z_{P}} -\tilde{F}^*_{,z} \, \tilde{G} \bigl|_{z_{T}}-\tilde{G}_{,z} \, \tilde{F}^* \bigl|_{z_{P}}+\tilde{G}_{,z} \, \tilde{F}^* \bigl|_{z_{T}}
    \right] ,
    \label{eq:re_hfg}
\end{align}
where $m_F^2$ and $m_G^{2}$ are eigenvalues of $\hat H$ corresponding to $\tilde F$ and $\tilde G$, respectively. 
 It should be noted that we obtain $(m^{2*}_F-m^2_F)(\tilde F,\tilde F)=0$ from Eq.~\eqref{eq:re_hfg},
 which implies that $m_F^2$ is real, $i.e.$ $m^{2*}_F-m^2_F=0$, because of $(\tilde F,\tilde F)>0$ for any non-trivial profile of $\tilde F_i$. 
One way to avoid the appearance of such non-Hermiticity is to focus on a specific boundary condition,
$d\tilde F/dz=\text{const.} \,\tilde F$, in which the extra boundary terms are canceled~\cite{Csaki:2000zn}. 
Such a boundary condition is obtained with infinite brane potential parameters of $\gamma_{\pm}\to \infty$, where Eq.~\eqref{varjp2} requires $\varphi\to 0$ and leads to $F'=2A' F$ at the boundaries.

Here, we give a way to recover the Hermiticity for finite brane potential parameters
by finding a new scalar product. 
The first step to this end is to write the boundary condition in Eq.~\eqref{varjp2}
in terms of only $\tilde F$ and  $d\tilde F/dz$.
At $z=z_{P}$, we obtain
\begin{align}
\tilde F(z)_{,z}=&\left(
\frac{4 \phi_P u \kappa^2 \left(\phi_{0,z}\right)^3 + \left(\phi_{0,z}\right)^2 \left(3 \gamma^2 A_{,z} + 9 e^{A(z)} \left(A_{,z}\right)^2 - 6 e^{A(z)} \left(m_F^2 - 2 A_{,zz}\right)\right)}{6 \phi_{0,z} \left(\left(\gamma^2 - e^{A(z)} A_{,z}\right) \phi_{0,z} - e^{A(z)} \phi_{0,zz}\right)}\right.  \notag\\
&\quad + \left.\frac{-3 \left(2 \gamma^2 - e^{A(z)} A_{,z}\right) \phi_{0,z} \phi_{0,zz} + 6 e^{A(z)} \left(\phi_{0,zz}\right)^2}{6 \phi_{0,z} \left(\left(\gamma^2 - e^{A(z)} A_{,z}\right) \phi_{0,z} - e^{A(z)} \phi_{0,zz}\right)}\right)\tilde F(z)\ ,
\label{z=0}\end{align}
while at $z=z_{T}$,
\begin{align}
\tilde F(z)_{,z}&=-\left( \frac{4 \phi_P u \kappa^2 \left(\phi_{0,z}\right)^3 - 3 e^{2\pi R u} \left(\phi_{0,z}\right)^2 \left(\gamma^2 A_{,z} - 3 e^{A(z)} \left(A_{,z}\right)^2 + 2 e^{A(z)} \left(m_F^2 - 2 A_{,zz}\right)\right)}{6 \phi_{0,z} \left(\gamma^2 + e^{A(z)} A_{,z}\right) \phi_{0,z} + e^{A(z)} \phi_{0,zz}}\right. \notag\\
&\quad \left.+ \frac{3 
\left(2 \gamma^2 + e^{A(z)} A_{,z}\right) \phi_{0,z} \phi_{0,zz} + 6 
e^{A(z)}
\left(\phi_{0,zz}\right)^2}{6 \phi_{0,z} \left(\gamma^2 + e^{A(z)} A_{,z}\right) \phi_{0,z} + e^{A(z)} \phi_{0,zz}}\right)e^{-2\pi R u} \tilde F(z)\ .
\label{z=z0}\end{align}
Here, $\varphi$ in Eq.~\eqref{varjp2} is rewritten in terms of $F$ and its derivative by Eq.~\eqref{bulkeq1}, and the second derivative for $F$ is eliminated with Eq.~\eqref{bulkeq2}. 
We can easily find that for another wavefunction $\tilde{G}$, the boundary conditions only differ by the eigenvalues, $m^2_F\to m^2_G$. Therefore, substituting Eqs.~\eqref{z=0} and~\eqref{z=z0} into Eq.~\eqref{eq:re_hfg}, we find
\begin{align}
(m_F^{2}-m_G^2)
\left[
\left(\tilde{F},\tilde{G}\right)
+
\frac{\tilde{F}^*\tilde{G}}{e^{-A(z)} \gamma^2 - A_{,z} - \frac{\phi_{0,zz}}{\phi_{0,z}}} \biggl|_{z=z_{P}}
+
\frac{\tilde{F}^*\tilde{G}}{e^{-A(z)} \gamma^2 + A_{,z} + \frac{\phi_{0,zz}}{\phi_{0,z}}} \biggl|_{z=z_{T}}
\right]=0\ .
\label{eq:n_p}
\end{align}
This motivates us to define a new scalar product,
\begin{equation}
    \left\{\tilde{F},\tilde{G}\right\}=\left(\tilde{F},\tilde{G}\right)+\frac{\tilde{F}\tilde{G}|_{z_{P}}}{e^{-A(z_{P})}\gamma^2-\alpha_{P}}+\frac{\tilde{F}\tilde{G}|_{z_{T}}}{e^{-A(z_{T})}\gamma^2-\alpha_{T}}\ ,
\end{equation}
where 
\begin{align}
    \alpha_{P}&=A_{,z} \Bigl|_{z=z_{P}} + \frac{\phi_{0,zz}}{\phi_{0,z}} \Bigl|_{z=z_{P}}\ ,\\[1ex]
    \alpha_{T}&=-A_{,z} \Bigl|_{z=z_{T}} - \frac{\phi_{0,zz}}{\phi_{0,z}} \Bigl|_{z=z_{T}}\ .
\end{align}
In terms of this new scalar product, Eq.~\eqref{eq:n_p} is expressed as
\begin{equation}
    \left(m_F^2-m_G^2\right)\left\{\tilde{F},\tilde{G}\right\}=0\ .
    \label{eq:r_n_p}
\end{equation}
For different eigenvalues, $m_F^2\neq m_G^2$, this relation gives the orthogonality of eigenfunctions,
$\left\{\tilde{F},\tilde{G}\right\}=0$.
Indeed, the Hermiticity is now recovered, $i.e.$ $0=\left(m_F^2-m_G^2\right)\left\{\tilde{F},\tilde{G}\right\}=\left\{\hat H\tilde{F},\tilde{G}\right\}-\left\{\tilde{F},\hat H\tilde{G}\right\}$.
Besides, the scalar product is consistently reduced to $(\tilde F,\tilde G)$
for the previously discussed limit of $\gamma_\pm \to \infty$.
Note that the new scalar product satisfies the commutative property $\{\tilde F,\tilde G\}=\{\tilde G,\tilde F\}$, the distributive property $\{\tilde F,(\tilde G+\tilde E)\}=\{\tilde F,\tilde G\}+\{\tilde F,\tilde E\}$, and the scalar multiplication property $\{\tilde F,\kappa\tilde G\}=\kappa \{\tilde F,\tilde G\}$ for $\kappa\in \mathbb{C}$, while the positivity $\{\tilde F,\tilde F\}\geq 0$ is broken in general.
The completeness condition for the new product is given as 
\begin{align}
    \sum_{n} \tilde F_n(z) \tilde F_n(z')
    =\delta(z-z')
    -\delta(z-z_{P}) \tilde \alpha_{P} \tilde F_{n}(z_{P}) \tilde F_{n}(z')
    -\delta(z-z_{T})  \tilde \alpha_{T} \tilde F_{n}(z_{T}) \tilde F_{n}(z')\ ,
\end{align}
where $\tilde \alpha_i=\frac{1}{e^{-A(z_{i})}\gamma^2-\alpha_{i}}$.
In appendix~\ref{app:kinetic_mixing}, we show that the newly defined scalar product appears in the kinetic terms of KK modes and makes it possible to remove the kinetic mixing between different modes.

\section{Perturbative analysis}\label{perturbative}

Let us now discuss the radion mass and wavefunction profile.
In the previous studies, the limit of $\gamma_\pm \to \infty$
in the brane-localized potentials of Eq.~\eqref{branepot} has been always assumed.
Here, we do not take this limit and clarify how a finite $\gamma$ modifies the radion mass and wavefunction profile.
However, in this case, it is difficult to find their analytical expressions. 
Then, we develop a series expansion in powers of $\gamma$ as well as the backreaction parameter defined by
$l\equiv\frac{\kappa \phi_P}{\sqrt{2}}$.%
\footnote{
We obtain a consistent massless radion mass and its profile for $l=0$.}

Substituting the background configuration of Eqs.~\eqref{phi0}, \eqref{A} into
the bulk equations~\eqref{bulkeq1}, \eqref{bulkeq2} and boundary conditions~\eqref{varjp2}, one obtains
\begin{eqnarray}
    &&F''+2\left(u-k+l^2 \frac{u}{3}e^{-2uy}\right)F' \nonumber \\
    &&\qquad +\left[m^2\exp\left(2ky+l^2\frac{e^{-2uy}}{3}\right)-l^2\frac{4u^2}{3}e^{-2uy}-4ku\right]F=0 \label{be1}\ ,\\[1ex]
    &&l^2\tilde{\varphi} + \frac{3}{2u}e^{uy}\left[F'-2\left(k-l^2\frac{u}{3}e^{-2uy}\right)F\right]=0\ , \label{be2} \\[1ex]
    &&\tilde{\varphi}'|_{0+\epsilon}=\gamma^2\tilde{\varphi}|_{0}-2uF|_{0}\ , \label{bc1}\\[1ex]
 &&\tilde{\varphi}'|_{\pi R-\epsilon} =-\gamma^2\tilde{\varphi}|_{\pi R} -2u e^{-u\pi R} F|_{\pi R}\ , \label{bc2}
\end{eqnarray}
with $\tilde{\varphi}\equiv \varphi/\phi_P$.
We now expand $F, \, \tilde{\varphi}, \, m^2$ in powers of $\frac{1}{\gamma^2}$, assuming a very large $\gamma$. 
In addition, to obtain an analytic solution, it is necessary to cope with the ``double exponential" term,
$\exp(l^2\frac{e^{-2uy}}{3})$, in Eq.~\eqref{be1}.
For this issue, we take the limit in which the backreaction parameter $l^2$ is small and
Taylor expand the exponential around $l^2=0$.
Therefore, $F, \, \tilde{\varphi}, \, m^2$ are expanded in terms of the two parameters $\frac{1}{\gamma^2}, l^2$ as
\begin{subequations}
    \begin{numcases}{}
    F(y)=\sum_{p,q}F_q^p(y)\cdot(l^{2})^p\cdot\left(\frac{1}{\gamma^2}\right)^q\label{F_expand}\ ,\\
    \tilde{\varphi}(y)=\sum_{p,q}\tilde{\varphi}_q^p(y)\cdot(l^{2})^p\cdot\left(\frac{1}{\gamma^2}\right)^q\label{phi_expand}\ ,\\
    m^2=\sum_{p,q}(m^2)_q^p(y)\cdot(l^{2})^p\cdot\left(\frac{1}{\gamma^2}\right)^q\equiv\sum_{p,q}M_q^p(y)\cdot(l^{2})^p\cdot\left(\frac{1}{\gamma^2}\right)^q\label{m_expand}\ ,
\end{numcases}
\end{subequations}
where the superscript $p$ and the subscript $q$ represent the orders of $l^2$ and $\frac{1}{\gamma^2}$, respectively.
We will use the notation $(p,q)$ with $p,q=0,1,2,\cdots $ to denote the equations of $p$-th order in $l^2$ and $q$-th order in $\frac{1}{\gamma^2}$.

We substitute the expansions into Eqs.~\eqref{be1}, \eqref{be2}, \eqref{bc1}, \eqref{bc2},
classify the equations in terms of the order of $(p,q)$ and perform the perturbative calculation
whose details are summarized in appendix~\ref{app:perturb_cal}.
The results of the radion mass and wavefunction up to the (1,1)th order are given by
\begin{eqnarray}
    m^2 &\approx& \frac{4u^2(2k+u)l^2}{3k}e^{-(2k+2u)\pi R}  \left(1-2\frac{2k+u}{\gamma^2} \right)
    +\mathcal{O} \left(l^2,{1}/{\gamma^2} \right) ,\label{m^2} \\[1ex]
    F(y) &\approx&  e^{2ky}+\frac{(k-u)l^2}{3k}e^{(2k-2u)y}
    -\left(1-\frac{4k+2u}{\gamma^2}\right)\frac{u^2l^2}{3k^2}e^{-(2k+2u)\pi R}e^{4ky}\notag\\
    &&+\left[\frac{u^2l^2}{3k(u+k)}+\frac{4u^2(2k+u)l^2}{3k(k+u)\gamma^2}e^{-(2k+2u)\pi R}\right]e^{-2uy}+\mathcal{O}\left(l^2,{1}/{\gamma^2} \right) .\label{wave}
\end{eqnarray}
Note that the radion mass vanishes in the limit of $l^2\to 0$.
This means that the backreaction of the scalar field on the metric is necessary
for a nonzero radion mass.
In the limit of $\gamma^2 \to \infty$, the radion mass reduces to $\frac{4u^2(2k+u) l^2}{3k} e^{-(2k+2u)\pi R}$,
which equals to the result derived in ref.~\cite{Csaki:2000zn}.
As the value of $\gamma^2$ decreases, the radion mass decreases.
We find a lower bound on $\gamma^2$ in which the perturbative calculation works:
\begin{equation}
    \gamma^2 > 4k+2u \ .
\label{eq:gamma_bound}
\end{equation} 
As shown in appendix~\ref{app:kinetic_mixing}, the same condition is obtained by computing
the coefficient of the radion kinetic term.

Eq.~\eqref{eq:gamma_bound} gives the leading order approximation of this lower bound of gamma $\gamma_c^2 = \gamma_0^2 = 4k+2u$,
where the radion becomes tachyonic as $\gamma^2 = \gamma_c^2$.
In addition, 
to show that $\gamma_0^2$ is a good approximation of $\gamma_c^2$,
we also evaluate $l^2$ order correction $\gamma_1^2$ so that $\gamma^2_c = \gamma_0^2 + l^2 \gamma^2_1$ and argue that higher $l^2$ order corrections to $\gamma_c^2$ are indeed negligible as the small backreaction limit $l^2 \ll 1$ is still valid.
To do so, we further computed the higher order corrections to the radion mass  $(m^2)^2_0$ and $(m^2)^2_1$ corresponding to $(l^4, 1/\gamma^0)$ and $(l^4, 1/\gamma^2)$ order respectively:

\begin{equation}
    \dfrac{(m^2)^2_0}{(m^2)^1_0} \approx -\dfrac{3k^3 + k^2 u - 2ku^2 - u^3}{3k^2(3k+u)} e^{-2u\pi R},\label{m20}
\end{equation}

\begin{equation}
    \dfrac{(m^2)^2_1}{(m^2)^2_0} \approx - (4k+2u) \dfrac{3k^4 + 7k^3 u + 2k^2 u^2 - 4ku^3 - 2u^4}{3k^4 + 4k^3 u - k^2 u^2 - 3k u^3 - u^4}.\label{m21}
\end{equation}
The above results are approximated by assuming the warp factor $e^{k \pi R} \sim M_{\text{Pl}}/M_{\text{EW}} \gg 1$ and $0< u \lesssim k$ for the parameter $u$ from the superpotential (to be more precise, in numerical analysis we take $k=37u$),
so that only those dominate exponential terms need to be kept.
Here $u>0$ guarantees the scalar bulk mass $\partial^2 V(\phi)/\partial \phi^2$ is not tachyonic,
and $u\lesssim k$ is taken to avoid introducing larger hierarchy than $e^{k \pi R} \sim M_{\text{Pl}}/M_{\text{EW}}$.
According to Eq.~\eqref{m_expand}, $(m^2)^2_0$ and $(m^2)^2_1$ contribute to $\gamma^2_1$ as
\begin{equation}
    \dfrac{\gamma_1^2}{\gamma_0^2} \approx -\dfrac{(m^2)^2_0}{(m^2)^1_0} \left(1+ 
    \dfrac{(m^2)^2_1}{(m^2)^2_0}/\gamma_0^2
    \right)
    \approx  -\left(\dfrac{u}{3k} + \mathcal O(u^2/k^2))\right)e^{-2u\pi R}.\label{gamma corr}
\end{equation}
With the moderate parameter choice $0< u \lesssim k$ (where $k\pi R \sim \mathcal O(10)$) above, one can estimate $|\gamma^2_1/\gamma^2_0| < 10^{-2}$ which means it barely affects $\gamma^2_c = \gamma_0^2 + l^2 \gamma^2_1$ taking account of the small backreaction limit $l^2\ll1$.
Other higher $l^2$ order corrections will be further suppressed by $l^2$ so that do not change the behavior of $\gamma_c^2$.

\section{Numerical analysis beyond perturbation}\label{numerical}

\begin{figure}[htbp]
    \centering
    \includegraphics[width=0.8\textwidth]{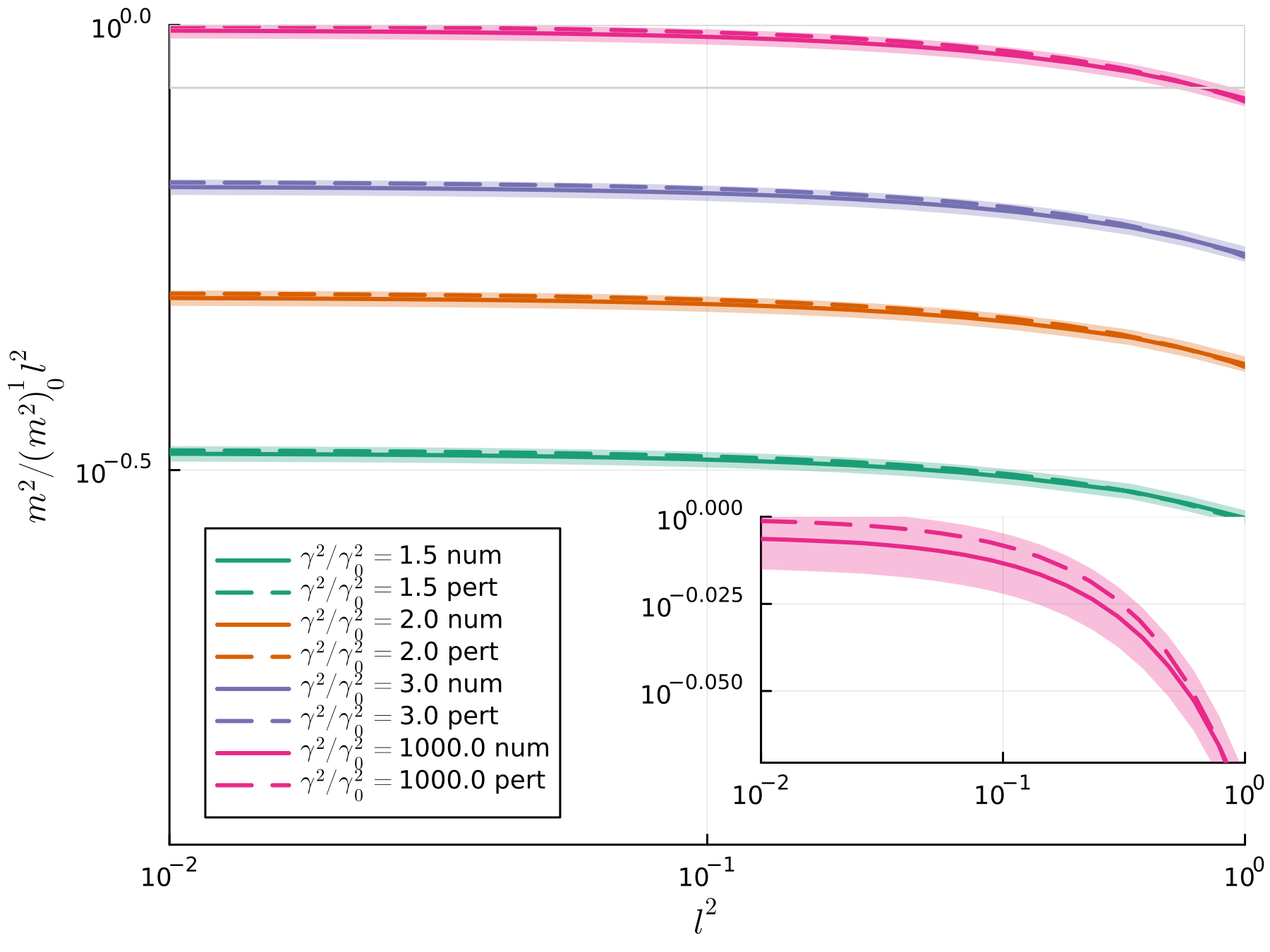}
    \caption{
        The radion mass with respect to the backreaction $l^2$.
        The radion mass $m$ is shown in unit of the leading order radion mass $l^2 (m^2)^1_0 = \frac{4u^2(2k+u)l^2}{3k}e^{-(2k+2u)\pi R}$. 
        We plot lines with various $\gamma^2$ in units of $\gamma_0^2 \equiv 4k+2u$.
        The dashed lines represent the perturbative results up to $(m^2)_1^2$ order and the solid lines are the numerical results.
        The region $l^2>1$ is not shown because 1). it is unnatural that scalar gets larger values than the 5d Planck scale 2). the negative quartic term in the scalar bulk potential $V$ becomes comparable to the mass term which may cause stability issues of the scalar profile.
        $u=37k$ and $k\pi R =3.7\pi$ are taken throughout the numerical calculation, which is consistent with the conditions $u>0$ for the positive scalar bulk mass $\partial^2 V/\partial \phi^2 >0$ and no larger hierarchy introduced $u \lesssim k$.
        The color shaded regions are 2\% relative error bands compared to the numerical results.
        The pink colored lines corresponding to the large gamma limit $\gamma^2\to \infty$ is magnified in the subplot in the bottom right. When $l^2$ gets larger, the decreasing behavior of radion masses relates to increasing importance of the negative quartic term in scalar bulk potential $V$ which makes the effective scalar bulk mass smaller.
       According to Eq.~\eqref{m20}, one can estimate the effect of varying $u$ within $0<u\ll k$ amounts to $1- \frac13 l^2e^{-2u\pi R}$ that just scales the changes of $m^2$ with respect to $l^2$.}
    \label{fig:mass_lm}
\end{figure}
\begin{figure}[htbp]
    \centering
    \includegraphics[width=0.8\textwidth]{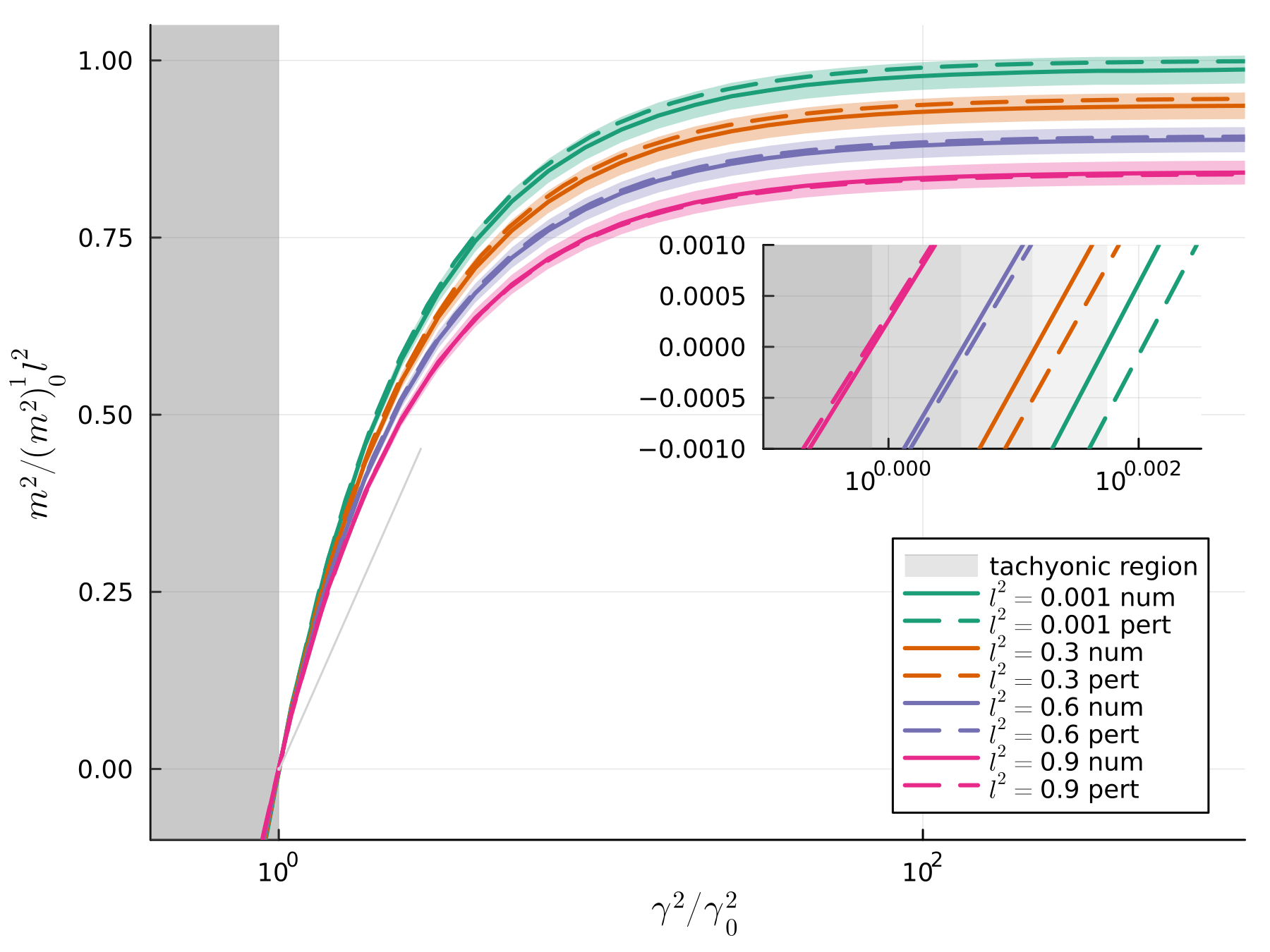}
    \caption{
        The $l^2$ normalized radion mass with respect to $\gamma^2$ in the unit of $\gamma_0^2$.
        The radion mass $m$ is shown in unit of the leading order radion mass $l^2 (m^2)^1_0 = \frac{4u^2(2k+u)l^2}{3k}e^{-(2k+2u)\pi R}$.  The dashed lines represent the perturbative result  up to $(m^2)_1^2$ order and the solid lines represent the numerical results
        with various $l^2$.
        $u=37k$ and $k\pi R =3.7\pi$ are taken throughout the numerical calculation, which is consistent with the conditions $u>0$ for the positive scalar bulk mass $\partial^2 V/\partial \phi^2 >0$ and no larger hierarchy introduced $u \lesssim k$.
        The color shaded regions are 2\% relative error bands compared to the numerical results.
        The gray shaded regions indicates the region where the radion mass-squared are negative, and their locations are all around $\gamma^2 = \gamma^2_0$, which is shown in the right subplot.
        The $l^2$ correction is relatively much smaller than the leading order results as expected by Eq.~\eqref{gamma corr}.
        The variation of $u$ has minor effect to $\gamma^2_c$ as is discussed around Eq.~\eqref{gamma corr}.
    }
    \label{fig:mass_gm}
\end{figure}

The perturbative approach assumes the small backreaction limit $l^2\ll 1$ and the large quadratic superpotential couplings $\gamma^2_\pm \gg 4k+2u$.
Here, we perform the numerical analysis of the radion mass without taking those limits.
The result is obtained by numerically solving $F$ and $\varphi$ by using the (modified) shooting method.
We vary the parameters $l^2$ and $\gamma^2$ to see the behavior of the radion mass.
The result shows
the tachyonic transition of the radion at $\gamma^2 \approx 4k+2u$
and the minor suppression of the radion mass from higher-order corrections of the backreaction,
which are consistent with the result of the perturbative approach.

The numerical analysis is performed in the following way, with the code available at \url{https://github.com/yardw/GoldbergerWiseRadionSpectra.jl}.
First, the metric component $F$ is solved from the bulk equation~\eqref{bulkeq2}. 
At this step, we note that the bulk equations and boundary conditions are already linearized,
so that the magnitudes of $F$ and $\varphi$ remain arbitrary 
until they are normalized with the proper scalar product discussed in section~\ref{non-hermiticity_section}.
Then, the scalar field $\varphi$ is solved from the bulk equation~\eqref{bulkeq1}.
This equation actually shows that the scalar field $\varphi$ can be read off from the metric component $F$ directly.
Consequently, the metric component $F$ is immune to the scalar fluctuation $\varphi$ on bulk (see Eq.~\eqref{bulkeq2}),
which allows us to put $\varphi$ aside when solving $F$ on bulk.
The boundary conditions~\eqref{varjp2} are checked at this step.
Here, we use a numerical approach based on the standard over/under shooting method for solving the boundary value problem.
The over/under shooting method is widely used in translating the boundary value problem to the initial value problem.
One starts from one end of the domain with a series of initial guesses
with different values of a field and its first-order derivative.
Usually, the degrees of freedom of the initial guesses are reduced
by one after checking a boundary condition at the end to be satisfied.
Then, the initial guess is evolved to the other end and another boundary condition is checked to be satisfied.
However, in our case, since the fields $\varphi$ and $F$ are coupled in the boundary conditions
and both fields are initially unknown,
a similar reduction of the degrees of freedom of the initial guesses does not occur by checking one boundary condition.
In addition, starting from one end leads to the amplification of the numerical error of $F$,
due to the exponential behavior of $F$ in the bulk.%
\footnote{{If the backreaction is very small and the true solution of $F$ is given by a single exponential function as seen from the analytical result, the error will be minimized by shooting near one end (the IR brane). In a generic parameter space, however, the true solution has a more complicated form so that shooting from one end is not a perfect way to minimize the error.}}
In order to address those issues, we start from the middle of the bulk and evolve the fields to both directions,
so that numerical errors on both sides are balanced to minimize the total error.
The initial guesses are parameterized by the radion mass-squared $m^2$ and the first-order derivative $F(y_0)'$
at $y_0 \in (0, \pi R)$.%
\footnote{{We numerically find $m^2$ and $F(y_0)'$ by solving the bulk field equations with boundary conditions while fixing the other parameters of $k$, $l$, $\gamma$, $etc$.}}
Toward the ends of the domain, we perform the over/under shooting method
to check that both boundary conditions are satisfied simultaneously.
In more details, 
in each point of the parameter space (made by fixed parameters of $k, l, \gamma, etc$), we consider a tangent space of $(m^2,F(y_0)')$.
Each tangent space contains two sets of vectors in which one (another) set satisfies the boundary condition at the UV (IR) brane within a certain precision.
For each set, we draw a fitting curve and look for the intersection of two fitting curves.%
\footnote{Both boundary conditions should be satisfied around the intersection.}
Once the intersection is found, 
new initial guesses are chosen at the neighborhood of the intersection
to improve the resolution.
The process is repeated until the error of the radion mass-squared $m^2$ is less than $10^{-6} M^2_{\rm IR}$
with $M_{\rm IR} \equiv \exp(-ky_T) M_{\rm Pl}$.

Fig.~\ref{fig:mass_lm} shows the result of the radion mass as a function of $l^2$.
We can see from the figure that the numerical results are in good agreement with the perturbative results
especially in the small $l^2$ limit.
When $l^2$ reaches $\mathcal O(1)$, the radion mass receives a minor suppression
from higher order corrections of the backreaction,
but the relative error of the perturbative result compared to the numerical result is still less than $15\%$.
Fig.~\ref{fig:mass_gm} describes the radion mass with respect to the quadratic superpotential coupling $\gamma^2$.
The radion mass decreases as $\gamma^2$ gets small, and finally 
the radion mode becomes tachyonic when $\gamma^2$ reaches $\gamma^2_c\sim4k+2u$,
which is consistent with the perturbative result.

\section{Conclusion}\label{conclusion}

We have studied radius stabilization in the RS model with a superpotential method
that takes into account the backreaction of the stabilizing scalar field in full generality
by keeping the effects of a finite $\gamma$ in the brane-localized potentials.
We developed a perturbative expansion in terms of the backreaction parameter and $1/\gamma^2$, and obtained analytical expressions for the radion mass and wavefunction up to the first order in the perturbation theory.
A numerical analysis without relying on the small backreaction and large $\gamma$ limits
was also performed, and it verified the validity of our perturbative results.
It was found that the radion mass decreases with the parameter $\gamma$, and below a critical value,
the mode becomes tachyonic.
This suggests that a minimum $\gamma$ is necessary for the stabilization of the extra dimension, denoted by $\gamma^2_c$. This is a reminiscence of earlier results in the literature concerning Karch-Randall geometries~\cite{Chacko_Fox_2001,Karch_Randall_2020, Mishra_2023}.

It was shown that a diverse range of the radion mass can be obtained by choosing an appropriate $\gamma$.
Moreover, we have addressed the issue of non-Hermiticity that arises due to the finite strip of extra dimension in the finite $\gamma$ case and defined an explicit general scalar product that restores the orthogonality in the KK decomposition and canonical kinetic terms, establishing the consistency of the method. 

It will be interesting to apply the methods of ref.~\cite{Konstandin:2010cd}
and analyze the gravitational waves from the phase transition in the present model for the parameter range
where we have shown a light radion mode appears
(for recent works on the light radion possibility, see ref.~\cite{Girmohanta:2023tdr} and references therein).
We demonstrated that the radion becomes tachyonic below a certain threshold $\gamma$,
and therefore, the extra dimension is destabilized.
It will be valuable to consider the holographic interpretation of such destabilization of the extra dimension
in terms of the 4D CFT.

\section*{Acknowledgements}

MS is supported by JSPS KAKENHI Grant Numbers JP22J00537.

\appendix

\section{Cancellation of kinetic mixing}
\label{app:kinetic_mixing}

Here, we check that the kinetic mixing between different KK modes is canceled due to the orthogonality of the new scalar product. 
As derived in the main text, the equation of motion for $\tilde{F}$ is given as
\begin{equation}
    \left(\square-\frac{d^2}{dz^2}\right)\tilde{F}(x^\mu,z)+\lambda(z)\tilde{F}(x^\mu,z) =0\ ,
\end{equation}
where $\square$ is the d'Amlembert operator in 4D spacetime and $\lambda(z)$ is given by Eq.~\eqref{lambda}.
We can construct the following bulk action leading to this bulk equation:
\begin{equation}
    \mathcal{S}_{bulk} = \int d^4x\int dz \left[\frac{1}{2}\left( \tilde{F}\square\tilde{F}+\left(\tilde{F}_{,z}\right)^2\right)+\frac{1}{2}\lambda(z) \tilde{F}^2 \right] \ ,
\end{equation}
Besides, the boundary conditions~\eqref{z=0} and \eqref{z=z0} indicate the boundary action,
\begin{align}
    \mathcal{S}_{brane} = & \int d^4 x \frac{1}{e^{-A}\gamma^2-\alpha_{P}}\left(\frac{1}{2}\tilde{F}\square\tilde{F}+\frac{1}{2}\lambda_{P}
    \tilde{F}^2\right) \biggr|_{z=z_{P}} \notag\ \\[1ex]
&+
    \int d^4 x \frac{1}{e^{-A}\gamma^2-\alpha_{T}}\left(\frac{1}{2}\tilde{F}\square\tilde{F}+\frac{1}{2}\lambda_{T} \tilde{F}^2\right) \biggr|_{z=z_{T}}\ ,
\end{align}
where $\lambda_{P, \, T}$ denote the coefficients of the r.h.s of Eqs.~\eqref{z=0}, \eqref{z=z0}
except for the eigenmasses,
\begin{align}
    \lambda_{P} =
&\frac{4 \phi_P u \kappa^2 \left(\phi_{0,z}\right)^3 + \left(\phi_{0,z}\right)^2 \left(3 \gamma^2 A_{,z} + 9 e^{A(z)} \left(A_{,z}\right)^2 + 12 e^{A(z_0)} A_{,zz}\right)}{6 \phi_{0,z} \left(\left(\gamma^2 - e^{A(z)} A_{,z}\right) \phi_{0,z} - e^{A(z)} \phi_{0,zz}\right)} \notag\ \\[1ex]
&\quad + \frac{-3 \left(2 \gamma^2 - e^{A(z)} A_{,z}\right) \phi_{0,z} \phi_{0,zz} + 6 e^{A(z)} \left(\phi_{0,zz}\right)^2}{6 \phi_{0,z} \left(\left(\gamma^2 - e^{A(z)} A_{,z}\right) \phi_{0,z} - e^{A(z)} \phi_{0,zz}\right)}\ ,\notag\\[2ex]
    \lambda_{T} =
&-e^{-2\pi R u}  \frac{4 \phi_P u \kappa^2 \left(\phi_{0,z}\right)^3 - 3 e^{2\pi R u} \left(\phi_{0,z}\right)^2 \left(\gamma^2 A_{,z} - 3 e^{A(z)} \left(A_{,z}\right)^2 - 4 e^{A(z)} A_{,zz}\right)}{6 \phi_{0,z} \left(\gamma^2 + e^{A(z)} A_{,z}\right) \phi_{0,z} + e^{A(z)} \phi_{0,zz}} \notag\\[1ex]
&\quad + \frac{3 
\left(2 \gamma^2 + e^{A(z)} A_{,z}\right) \phi_{0,z} \phi_{0,zz} + 6 
e^{A(z)}
\left(\phi_{0,zz}\right)^2}{6 \phi_{0,z} \left(\gamma^2 + e^{A(z)} A_{,z}\right) \phi_{0,z} + e^{A(z)} \phi_{0,zz}} \ .
\end{align}
The total action is $\mathcal{S} = \mathcal{S}_{bulk} + \mathcal{S}_{brane}$.

To check the orthogonality of the kinetic terms of modes in the 4D effective action concretely,
we replace $\tilde{F}$ with $\tilde{F}(x^\mu,z)=\tilde{F}_0(z)r_0(x^\mu)+\tilde{F}_1(z)r_1(x^\mu)+...$ where $...$ denotes the higher KK excitation modes, and integrate over the fifth dimension,
\begin{align}
    \mathcal{S}_{kinetic} =& \int d^4x\int dz \frac{1}{2} \tilde{F}\square\tilde{F}+\int d^4 x \frac{1}{e^{-A}\gamma^2-\alpha_{P}}\frac{1}{2}\tilde{F}\square\tilde{F} \biggr|_{z=z_{P}}
    +\int d^4 x \frac{1}{e^{-A}\gamma^2-\alpha_{T}}\frac{1}{2}\tilde{F}\square\tilde{F} \biggr|_{z=z_{T}}\notag\ \\
    =&\int d^4 x \, r_0\square r_1 \left\{\tilde{F}_0,\tilde{F}_1\right\} \nonumber \\
    &+\int d^4 x \, r_0\square r_0 \left[\int dz \tilde{F}_0^2+\frac{\tilde{F}_0^2}{e^{-A}\gamma^2-\alpha_{P}} \biggr|_{z=z_{P}}
    +\frac{\tilde{F}_0^2}{e^{-A}\gamma^2-\alpha_{T}} \biggr|_{z=z_{T}}
    \right]\notag\\
    &+\int d^4 x \, r_1\square r_1 \left[\int dz \tilde{F}_1^2+\frac{\tilde{F}_1^2}{e^{-A}\gamma^2-\alpha_{P}} \biggr|_{z=z_{P}}
    +\frac{\tilde{F}_1^2}{e^{-A}\gamma^2-\alpha_{T}} \biggr|_{z=z_{T}}\right]+...\ ,
\end{align}
where we explicitly write down the zero and first KK excitation modes.
Note that the cross term is proportional to the redefined scalar product and thus this should vanish by the orthogonality. We can also show the vanishing kinetic mixing with higher KK excitation modes.
By computing the coefficient of the kinetic term $r_0 \square r_0$ perturbatively, we find the condition that the $0$-th order term of $1/\gamma^2$ is larger than the 1-th order term gives $\gamma^2 >  4k+u$,
which is consistent with the result of Eq.~\eqref{eq:gamma_bound}.

\section{Details of perturbative calculation}
\label{app:perturb_cal}

We solve the bulk equations~\eqref{be1}, \eqref{be2} with the boundary conditions~\eqref{bc1}, \eqref{bc2} to
find the radion mass-squared and wavefunction by a perturbative expansion
in terms of the two parameters $l^2, 1/\gamma^2$.
The expansions~\eqref{F_expand}, \eqref{phi_expand}, \eqref{m_expand} lead to
groups of equations corresponding to different orders of $(p,q)$.
We will solve them order by order.

Let us start with the $(0,0)$th order equations,
\begin{subequations}
    \begin{numcases}{}
    {F_0^0}''+2(u-k){F_0^0}'-4kuF_0^0+M_0^0 e^{2ky}F_0^0=0\label{00a}\ ,\\[1ex]
    \label{00b}{F_0^0}'-2kF_0^0=0\ ,\\[1ex]
    \label{00c}\tilde{\varphi}_0^0|_i=0\ .
    \end{numcases}
\end{subequations}
Eq.~\eqref{00b} gives
\begin{equation}
F_0^0=C_0 e^{2ky}\ ,
\end{equation}
with $C_0$ being a mass dimension one constant.
We substitute this into Eq.~\eqref{00a} to get $M_0^0=0$.
The $(1,0)$th order equations are:
\begin{subequations}
    \begin{numcases}{}
    {F_0^1}''+2(u-k){F_0^1}'-4kuF_0^1+M_0^1 e^{2ky}F_0^0+\frac{2u}{3}e^{-2uy}({F_0^0}'-2uF_0^0)=0\label{10a}\ ,\\[1ex]
    \label{10b}\frac{2u}{3}\tilde{\varphi}_0^0=-e^{uy} \left[{F_0^1}'-2kF_0^1+\frac{2u}{3}e^{-2uy}F_0^0 \right]\ ,\\[1ex]
    \label{10c}\tilde{\varphi}_0^1|_i=0\ .
    \end{numcases}
\end{subequations}
Substituting $F_0^0$ into Eq.~\eqref{10a} and Eq.~\eqref{10b}, we can solve for $F_0^1$ and $\tilde{\varphi}_0^0$ as
\begin{subequations}
    \begin{numcases}{}
    F_0^1 = \frac{C_0(k-u)}{3k}e^{(2k-2u)y}-\frac{C_0 M_0^1}{4k(2k+u)}e^{4ky}+D_0^1 e^{-2uy}+D_0^2 e^{2ky}\label{F10}\ ,\\[1ex]
    \label{phi00}\tilde{\varphi}_0^0 = -\frac{3}{2u}e^{uy} \left[\frac{2C_0 u^2}{3k}e^{(2k-2u)y}-\frac{C_0 M_0^1}{2(2k+u)}e^{4ky}-D_0^1(2u+2k)e^{-2uy} \right]\ ,
    \end{numcases}
\end{subequations}
where $D_0^1$ and $D_0^2$ are two integration constants. 
Then, substituting the expression for $\tilde{\varphi}_0^0$ into the boundary condition~\eqref{00c}, one obtains
\begin{subequations}
    \begin{numcases}{}
    M_0^1 = \frac{4(2k+u)u^2}{3k}\frac{1-e^{2k\pi R}}{1-e^{(4k+2u)\pi R}}\approx \frac{4(2k+u)u^2}{3k} e^{-(2k+2u)\pi R}\label{M10}\ ,\\[1ex]
    \label{D10}D_0^1 = \frac{C_0 u^2}{3k(u+k)}\frac{e^{2k\pi R}-e^{(4k+2u)\pi R}}{1-e^{(4k+2u)\pi R}}\ .
    \end{numcases}
\end{subequations}
The radion mass at the $(1,0)$th order in Eq.~\eqref{M10} equals the result presented in ref.~\cite{Csaki:2000zn}. 

Next, we study the first order in $\frac{1}{\gamma^2}$.
The (0,1)th order equations are:
\begin{subequations}
    \begin{numcases}{}
    {F_1^0}''+2(u-k){F_1^0}'-4kuF_1^0+M_1^0 e^{2ky}F_0^0 =0\label{01a}\ ,\\[1ex]
    \label{01b}{F_1^0}'-2kF_1^0=0\ ,\\[1ex]
    \label{01c}({\tilde{\varphi}_0^0})'|_{0+\epsilon}=\tilde{\varphi}_1^0|_0-2uF_0^0|_0\ ,\\[1ex]
    \label{01d}({\tilde{\varphi}_0^0})'|_{\pi R-\epsilon}=-\tilde{\varphi}_1^0|_{\pi R}-2u e^{-u \pi R} F_0^0 |_{\pi R}\ .
    \end{numcases}
\end{subequations}
The two bulk equations give
\begin{subequations}
    \begin{numcases}{}
    F_1^0=C_1 e^{2ky}\ ,\\[1ex]
    M_1^0=0\ ,
    \end{numcases}
\end{subequations}
with $C_1$ being a mass dimension one constant.
Then, the (1,1)th order equations are:
\begin{subequations}
    \begin{numcases}{}
    {F_1^1}''+2(u-k){F_1^1}'-4kuF_1^1+e^{2ky}(M_1^1 F_0^0+M_0^1 F_1^0)+\frac{2u}{3}e^{-2uy}({F_1^0}'-2uF_1^0)
    =0\label{11a}\ ,\\[1ex]
    \label{11b}\frac{2u}{3}\tilde{\varphi}_1^0=-e^{uy} \left[{F_1^1}'-2kF_1^1+\frac{2u}{3}e^{-2uy}F_1^0 \right]\ ,\\[1ex]
    \label{11c}({\tilde{\varphi}_0^1})'|_{0+\epsilon}=\tilde{\varphi}_1^1|_0-2uF_0^1|_0\ ,\\[1ex]
    \label{11d}({\tilde{\varphi}_0^1})'|_{\pi R-\epsilon}=-\tilde{\varphi}_1^1|_{\pi R}-2u e^{-u\pi R} F_0^1 |_{\pi R}\ .
    \end{numcases}
\end{subequations}
The bulk equations~\eqref{11a}, \eqref{11b} lead to general solutions,
\begin{subequations}
    \begin{numcases}{}
    F_1^1 = \frac{C_1(k-u)}{3k}e^{(2k-2u)y}-\frac{C_1 M_0^1+C_0 M_1^1}{4k(2k+u)}e^{4ky}+D_1^1 e^{-2uy}+D_1^2 e^{2ky}\label{F11}\ ,\\[1ex]
    \label{phi01}
    \tilde{\varphi}_1^0 = -\frac{3}{2u}e^{uy}\left[\frac{2C_1 u^2}{3k}e^{(2k-2u)y}-\frac{C_1 M_0^1+C_0 M_1^1}{2(2k+u)}e^{4ky}-D_1^1(2u+2k)e^{-2uy}\right]\ .
    \end{numcases}
\end{subequations}
Using the boundary conditions~\eqref{01c}, \eqref{01d}, we find
\begin{subequations}
    \begin{numcases}{}
    M_1^1 = \frac{8u^2(2k+u)^2}{3k}\frac{(1-e^{2k\pi R})(1+e^{(4k+2u)\pi R})}{(1-e^{(4k+2u)\pi R})^2}\label{M11}\ ,\\[1ex]
    \label{D11}D_1^1 = \frac{C_1 u^2}{3k(k+u)}\frac{e^{2k}-e^{4k+2u}}{1-e^{4k+2u}}+\frac{4C_0 u^2(2k+u)}{3k(k+u)}\frac{e^{(4k+2u)\pi R}(e^{2k\pi R}-1)}{(1-e^{(4k+2u)\pi R})^2}\ .
    \end{numcases}
\end{subequations}
Therefore, we have the eigenmass and wavefunction for the radion up to the first order in both $l^2$ and $\frac{1}{\gamma^2}$:
\begin{align}
    m^2 &= \frac{4u^2(2k+u)l^2}{3k}\frac{1-e^{2k\pi R}}{1-e^{(4k+2u)\pi R}}  \left(1+2\frac{2k+u}{\gamma^2}\frac{1+e^{(4k+2u)\pi R}}{1-e^{(4k+2u)\pi R}} \right)+\mathcal{O} \left(l^2, {1}/{\gamma^2} \right)\notag\\[1ex]
    &\approx \frac{4u^2(2k+u)l^2}{3k}e^{-(2k+2u)\pi R}  \left(1-2\frac{2k+u}{\gamma^2} \right)+\mathcal{O} \left(l^2, {1}/{\gamma^2} \right) ,\\[2ex]
    F(y) &= \left(C_0+\frac{C_1}{\gamma^2}+l^2 D_0^2+\frac{l^2}{\gamma^2}D_1^2\right)e^{2ky}+\left(C_0+\frac{C_1}{\gamma^2}\right)\frac{(u-k)l^2}{3k}e^{(2k-2u)y}\notag \\[1ex]
    &-\left(C_0+\frac{1}{\gamma^2} \left[C_1+2C_0(2k+u)\frac{1+e^{(4k+2u)\pi R}}{1-e^{(4k+2u)\pi R}} \right]\right)\frac{u^2l^2}{3k^2}\frac{1-e^{2k\pi R}}{1-e^{(4k+2u)\pi R}}e^{4ky}\notag\\[1ex]
    &+\left[\left(C_0+\frac{C_1}{\gamma^2}\right)\frac{u^2l^2}{3k(u+k)}\frac{e^{2k\pi R}-e^{(4k+2u)\pi R}}{1-e^{(4k+2u)\pi R}}+\frac{C_0}{\gamma^2}\frac{4u^2(2k+u)l^2}{3k(k+u)}\frac{e^{(4k+2u)\pi R}(e^{2k\pi R}-1)}{(1-e^{(4k+2u)\pi R})^2}\right]e^{-2uy}\notag\\[1ex]
    &+ \mathcal{O} \left(l^2, {1}/{\gamma^2} \right) .\label{wave1}
\end{align}
The wavefunction contains many integral constants,
which seem to make it undetermined, but in our perturbative method,
they can be eliminated by redefinition of constants. For example, the redefinition,
\begin{equation}
    C_0+\frac{C_1}{\gamma^2}=C\ ,
\end{equation}
gives $C_0 = C - \frac{C_1}{\gamma^2}$. Substituting this into the original wavefunction, one obtains
\begin{align}
    F(y) =& \left(C+l^2 D_0^2+\frac{l^2}{\gamma^2}D_1^2\right)e^{2ky}+C\frac{(k-u)l^2}{3k}e^{(2k-2u)y}\notag\\[1ex]
    -&\left(C+\left(\frac{C}{\gamma^2}-\frac{C_1}{\gamma^4}\right)(4k+2u)\frac{1+e^{(4k+2u)\pi R}}{1-e^{(4k+2u)\pi R}}\right)\frac{u^2l^2}{3k^2}\frac{1-e^{2k\pi R}}{1-e^{(4k+2u)\pi R}}e^{4ky}\notag\\[1ex]
    +&\left[C\frac{u^2l^2}{3k(u+k)}\frac{e^{2k\pi R}-e^{(4k+2u)\pi R}}{1-e^{(4k+2u)\pi R}}+\left(\frac{C}{\gamma^2}-\frac{C_1}{\gamma^4}\right)\frac{4u^2(2k+u)l^2}{3k(k+u)}\frac{e^{(4k+2u)\pi R}(e^{2k\pi R}-1)}{(1-e^{(4k+2u)\pi R})^2}\right]e^{-2uy}\notag\\[1ex]
    +&\mathcal{O} \left(l^2, {1}/{\gamma^2} \right) .
\end{align}
Since the $C_1$ terms are both of order $\frac{1}{\gamma^4}$, they should be neglected.
The same argument can be done for $D_0^2$ and $D_0^4$, and we finally get
\begin{align}
    F(y) =& \, e^{2ky}+\frac{(k-u)l^2}{3k}e^{(2k-2u)y}
    -\left(1+\frac{4k+2u}{\gamma^2}\frac{1+e^{(4k+2u)\pi R}}{1-e^{(4k+2u)\pi R}}\right)\frac{u^2l^2}{3k^2}\frac{1-e^{2k\pi R}}{1-e^{(4k+2u)\pi R}}e^{4ky}\notag\\[1ex]
    +&\left[\frac{u^2l^2}{3k(u+k)}\frac{e^{2k\pi R}-e^{(4k+2u)\pi R}}{1-e^{(4k+2u)\pi R}}+\frac{4u^2(2k+u)l^2}{3k(k+u)\gamma^2}\frac{e^{(4k+2u)\pi R}(e^{2k\pi R}-1)}{(1-e^{(4k+2u)\pi R})^2}\right]e^{-2uy}+\mathcal{O}\left(l^2, {1}/{\gamma^2} \right)\notag \\[1.5ex]
    \approx & \, e^{2ky}+\frac{(k-u)l^2}{3k}e^{(2k-2u)y}
    -\left(1-\frac{4k+2u}{\gamma^2}\right)\frac{u^2l^2}{3k^2}e^{-(2k+2u)\pi R}e^{4ky}\notag\\[1ex]
    +&\left[\frac{u^2l^2}{3k(u+k)}+\frac{4u^2(2k+u)l^2}{3k(k+u)\gamma^2}e^{-(2k+2u)\pi R}\right]e^{-2uy}+\mathcal{O}\left(l^2, {1}/{\gamma^2} \right) ,
\end{align}
where we have set the overall coefficient $C$ to 1 for simplicity.
\bibliographystyle{jhep}
\bibliography{bib}

\end{document}